\documentclass[article,twocolumn,9pt]{IEEEtran}
\makeatletter
\def\ps@headings{%
\def\@oddhead{\mbox{}\scriptsize\rightmark \hfil \thepage}%
\def\@evenhead{\scriptsize\thepage \hfil \leftmark\mbox{}}%
\def\@oddfoot{}%
\def\@evenfoot{}}
\makeatother 

\usepackage{graphicx}
\usepackage{float}
\usepackage{amsfonts}
\usepackage{amsmath}
\usepackage{amssymb}
\usepackage{cite}
\usepackage{bm}
\usepackage{mathrsfs}
\usepackage{epsfig}
\usepackage{algorithm}
\usepackage{algorithmic}
\usepackage{setspace}
\usepackage{setspace}

\begin{document}
\title{Cooperative Sparsity Pattern Recovery in Distributed  Networks Via Distributed-OMP}
\author{\authorblockA{Thakshila Wimalajeewa and Pramod K.
Varshney}\\
Department of Electrical Engineering  and Computer Science,\\
Syracuse University, Syracuse, NY 13244\\
Email: twwewelw@syr.edu, varshney@syr.edu}

\maketitle \thispagestyle{empty}

\begin{abstract}
In this paper, we consider the problem of collaboratively estimating the  sparsity pattern of a  sparse signal with multiple measurement data in distributed networks. We assume that each node makes Compressive Sensing (CS) based measurements via random projections regarding the same sparse signal.  We propose a distributed greedy algorithm based on Orthogonal Matching Pursuit (OMP), in which the  sparse support is estimated iteratively while fusing indices  estimated at distributed nodes. In the proposed distributed framework, each node has to perform  less  number of iterations of OMP compared to the sparsity index of the sparse signal. Thus,  with  each node having a very small number of compressive measurements, a significant performance gain in support recovery  is achieved via the proposed collaborative scheme compared to the case where each node estimates  the sparsity pattern independently  and then  fusion is performed  to get a global estimate. We further extend the algorithm to estimate the sparsity pattern in a binary hypothesis testing framework, where the algorithm first detects the presence of a sparse signal collaborating among nodes with a fewer number of iterations of OMP and then increases the number of iterations to estimate the sparsity pattern only if the signal is detected.
\end{abstract}


\section{Introduction}

In the Compressive Sensing (CS) framework, a small collection of
linear random projections of a sparse signal contains sufficient information
for complete signal recovery \cite{donoho1,candes3, baraniuk1}.
There is a considerable amount of work on the development of
computationally efficient and tractable algorithms to recover sparse signals from CS based measurements obtained via random projections, for example in   \cite{candes4,candes5,Romberg1,Laska2,tropp1,Figueiredo1,Blumensath1,Blumensath2,Needell1}.
However, there are several
signal processing applications
where complete signal recovery is not necessary. For example, in applications such as source localization in sensor networks \cite{Malioutov1,Cevher1}, estimation of frequency band locations in cognitive radio networks \cite{Tian1}, localization of  neural electrical activities from a huge number of potential locations in magnetoencephalography (MEG) and electroencephalography
(EEG) for  medical imaging applications \cite{Jin2,Baillet1,Wipf1},
sparse approximation \cite{Natarajan1}, subset selection in linear regression \cite{Miller1,Larsson1},
and signal denoising \cite{SSChen1}, it is only necessary to estimate the locations of the significant coefficients of a sparse signal.


\subsection{Related work and our contribution}
The problem of sparsity pattern estimation of sparse signals has been addressed by several authors in the literature in the context of CS \cite{wain1,wain2,wang5,Fletcher1,Akcakaya1,Reeves1, tang1,goyal1,thakshilaj4}.  These studies focus on CS based sparsity pattern recovery  with a single measurement vector. The achievable performance can be improved by having multiple observation vectors. Further, in distributed networks including sensor networks and cooperative cognitive radio networks, multiple measurements appear quite naturally since multiple nodes make observations regarding the same phenomenon of interest (PoI). Extensions of sparse signal recovery with multiple measurement vectors are considered  in \cite{Tropp2,Baron1}  assuming  that the multiple measurements are available at a single location to perform the desired task. In such centralized settings, each node in the network has to transmit its observations along with the elements of the measurement matrix to solve the sparse signal  recovery problem at the central fusion center.   However, when performing inference tasks based on the observations collected at different nodes in distributed networks, the trade-off between the resource constraints (e.g. node power and communication bandwidth) and the achievable performance gain is a core issue to be addressed in many practical networks. Thus, distributed approaches for sparsity pattern recovery are desirable in many practical distributed networks  where each node has limited  computational and communication power/bandwidth.  The problem of distributed sparsity pattern  recovery is considered recently  in the context of cognitive radio networks in \cite{ling1,Zeng1}. There, decentralized consensus based algorithms for support recovery of sparse signals were  proposed when  each cognitive radio makes CS based measurements   in  cooperative cognitive radio networks.

The  work in this paper focuses on further reducing the computational and communication burden at individual nodes in a distributed network while performing sparsity pattern recovery when each node in the network observes a noisy version of the same sparse signal.  In contrast to transmitting raw observations along with measurement matrices to a central fusion center to perform  centralized  sparsity pattern recovery, we consider the case where each node tries to find an estimate of the sparsity pattern by collaboration and fusion. More specifically, we develop a greedy algorithm based on Orthogonal Matching Pursuit (OMP) where the indices  of the sparse support are iteratively identified while fusing the estimated indices at each iteration. We show that, in  the proposed distributed algorithm, each node has to perform  less number of iterations compared to the sparsity index of the sparse signal to reliably estimate the sparsity pattern (in the centralized  OMP algorithm,  at least $K$  iterations are required for sparsity pattern recovery where $K$ is the sparsity index of the sparse signal). Moreover, in the proposed algorithm, each node transmits only  one index at each iteration.   Further, we extend the results and develop  a joint algorithm to both detect the sparse signal and to perform the sparsity  pattern recovery only if the sparse signal is detected.



\section{Problem Formulation and Background}\label{sparsity}
Consider a set of $L$ distributed nodes making  noisy measurements on a sparse signal.  We assume that each sensor node obtains a  $M (< N)$-length measurement vector  $\mathbf y_l$  via CS based linear random projections.
The measurement vector at $l$-th sensor node is  given by,
\begin{eqnarray}
\mathbf y_l&=&\mathbf A_l \mathbf s + \mathbf v_l;\label{obs_1}
\end{eqnarray}
for $l=0,1,\cdots,L-1$. $\mathbf s$ is the $N\times 1$ sparse signal and  $\mathbf A_l$ is the $M\times N$ random projection  matrix at the $l$-th  node.
The noise vector $\mathbf v_l$ at the $l$-th sensor node is assumed to be iid Gaussian with zero mean vector and the covariance matrix $\sigma_{v}^2 \mathbf I_M$ where $\mathbf I_M$ is the $M\times M $ identity matrix.

When the signal $\mathbf s$ is sparse  in the basis $\Psi$ such that $\mathbf s = \Psi \boldsymbol\beta$ where $\boldsymbol\beta$ contains only $K << N$ number of significant coefficients, it has been shown that the randomized lower dimensional projections of the form (\ref{obs_1}) can capture significant  information of the sparse signal $\mathbf s$. Assume  that one has to estimate the sparsity pattern of a sparse signal which will give important information in many applications. For example, if the signal is sparse in Fourier  domain, the locations of  non zero coefficients give the locations of significant harmonics which is important  in spectrum sensing in the wideband regime.

Let $\mathcal U$ be the support of the sparse coefficient vector $\boldsymbol\beta$ which is defined as,
$
\mathcal U := \{i\in \{1,2,\cdots,N\}~|~ \boldsymbol\beta (i) \neq 0 \}
$
where $\boldsymbol\beta (i)$ is the $i$-th element of $\boldsymbol\beta$ for $i=0,1,\cdots, N-1$. Then we have $K=n(\mathcal U)$ where $n(\mathcal S)$ denotes the cardinality of  the set $\mathcal S$. Further, let $\mathbf b$ be a $N$-length vector which contains binary elements:
i.e.
\begin{eqnarray*}
\mathbf b(i) = \left\{
\begin{array}{ccc}
1~ & if \boldsymbol\beta(i) \neq 0\\
0 ~& otherwise
\end{array}\right.
\end{eqnarray*}
for $i=0,1,\cdots, N-1$ and $\hat{\mathbf b}$ be the estimated binary support vector.

%
%
Sparsity pattern recovery with a single measurement vector can be performed via  a CS reconstruction algorithm such as regularized least square algorithm (Lasso) \cite{donoho1,Tibshirani1} or OMP \cite{tropp1}.
When multiple measurement vectors as in (\ref{obs_1}) are collected  at a centralized location, the support of the sparse signal can be estimated, for example,  by solving the  simultaneous OMP algorithm as given in \cite{Tropp2}\cite{Baron1} in which the support can be directly estimated iteratively without reconstructing the complete sparse signal $\boldsymbol\beta$.

\section{Sparsity pattern recovery with multiple measurement vectors: Distributed Algorithm}\label{omp_dist}
To implement a centralized sparsity pattern recovery algorithm based on the measurements collected at distributed nodes in a distributed network,  it is required that each node transmits its  $M$-length observation vector along with the elements of the measurement matrix  to a central fusion center.  Since transmitting all the information to a fusion center is not desirable in power and bandwidth  constrained communication  networks,  we consider a distributed scheme in which, sparsity pattern estimation is performed via collaboration among nodes with less communication and with  each node estimating the sparsity pattern. In \cite{ling1,Zeng1}, several consensus based distributed schemes are proposed to estimate the support of sparse signals based on Lasso. These schemes estimate the full support set  at each node and then perform fusion via several collaboration schemes. However, due to computational complexity, performing Lasso at power constrained sensor nodes may  not be desirable.

OMP is a greedy approach in which at each iteration, the location of one non zero coefficient or a  column of $\Theta=\mathbf A \Psi$ that participates in the measurement vector $\mathbf y$ is identified. More specifically, at each iteration, it picks   the column of $\Theta$ which is most correlated with the remaining part of $\mathbf y$. Then the selected column's  contribution is subtracted from $\mathbf y$ and iterations  on the residual are carried out. If we consider one sensor node, at each iteration, there are $N-K$ possible incorrect indices that can be selected by the OMP algorithm. The probability of selecting  an incorrect index at each iteration increases as the number of CS measurements at each node ($M$) decreases. It has been shown that the OMP algorithm requires more measurements for signal reconstruction compared to optimization based (e.g. Basis Pursuit (BP)) algorithms. However, due to limitation of processing power at each node in many practical distributed networks, it is desired to have multiple nodes each processing a small number of measurements. Since all the nodes in the network observe the same sparse signal, it is highly  likely that the estimate of the index at multiple nodes is the  same at a given iteration of OMP.  To reduce the error in incorrectly selecting an index at each iteration of OMP with less number of compressive measurements, we propose to fuse the indices  estimated by each node during a given iteration by collaboration  among the  nodes in the network.   It is worth mentioning that, in the proposed approach, a node in the network may find a set of indices (instead of one index as in the conventional OMP) that  correspond to non zero coefficients  via collaboration, especially when the number of distributed  nodes is  closer to or  greater than the sparsity index. Thus the proposed  algorithm  can be terminated with   less number of iterations compared to $K$ at each node. It should be further noted that, in  the proposed OMP based algorithm, at each iteration, only one index needs to be transmitted by each node for  fusion, thus reducing communication cost compared to that with distributed Lasso versions proposed in \cite{ling1,Zeng1}. We dub the proposed algorithm as `Distributed and collaborative  OMP (DC-OMP)'.

\subsubsection*{Distributed and collaborative OMP for sparse support estimation}
Define $\mathcal M_l$ to be the set containing the neighboring nodes of the $l$-th  node (including itself).
As defined in Subsection \ref{sparsity}, let $\mathcal U$ be the support set which contains the indices of non zero coefficients of the sparse signal and  $\hat{\mathcal U_l}$ be the estimated support set at the $l$-th node.
Further, let $\Theta_{l} = \mathbf A_l \Psi$ and $\theta_{l,i}$ be the $i$-th column of the matrix $\Theta_l$. $\Theta_l(\mathcal A)$ denotes the submatrix of $\Theta_l$ which has columns of $\Theta_l $ corresponding  to the elements in the set $\mathcal A$ for  $\mathcal A \subset \{0,1,\cdots,N-1\}$.  $|.|$ denotes the absolute value while $||.||_p$ denotes the $l_p$ norm. Further, $n(\mathcal S)$ denotes the cardinality of the set $\mathcal S$ as defined before.

In the proposed DC-OMP algorithm which is stated   in Algorithm \ref{algo1},  once the $l$-th node finds an index $\lambda_l(t)$ (which corresponds to the column that is most correlated with the remaining part of $\mathbf y_l$) at iteration $t$ by performing step 2 in Algorithm \ref{algo1}, it is exchanged among the neighborhood $\mathcal M_l$. Subsequently, each node will have the index that the nodes in its neighborhood obtained from step 2. By fusion, each node selects a set of indices (from $n(\mathcal M_l)$ number of indices)  such that most of the nodes agree upon (more details of this fusion are provided  in Subsection \ref{step_3}). Note that, in this step several indices can be selected and thus, the number of iterations required to estimate the support that each node has to perform can be less than the sparsity index $K$.

\begin{algorithm}
At $l$-th node:
\begin{enumerate}
\item Initialize $t=1$, $\hat{\mathcal U}_l(0) = \emptyset $, residual vector $\mathbf r_{l,0} = \mathbf y_l$
\item Find the index $\lambda_l(t)$ such that
\begin{eqnarray*}
\lambda_l(t) = \underset{\omega = 1,\cdots,N}{arg~ max} ~ \frac{|<\mathbf r_{l,t-1}, \theta_{l,\omega}>|}{||\mathbf r_{l,t-1}||_2}
\end{eqnarray*}
 \item Update the index set $\lambda_l^*(t)$ via local communication: $\lambda_l^*(t) = f_l(\lambda_l(t),\{\lambda_{i}(t)\}, i\in \mathcal M_l)$, as discussed in subsection \ref{step_3}
 \item  Set $\hat{\mathcal U}_l(t) = \hat{\mathcal U}_l(t-1) \cup \{\lambda_l^* (t)\}$ and $l_t = n(\hat{\mathcal U}_l(t))$
\item  Compute the projection operator $\mathbf P_l(t) = \Theta_l(\hat{\mathcal U}_l(t)) \left(\Theta_l(\hat{\mathcal U}_l(t)) ^T \Theta_l(\hat{\mathcal U}_l(t)) \right)^{-1} \Theta_l(\hat{\mathcal U}_l(t)) ^T$. Update the residual vector:  $\mathbf r_{l,t} = (\mathbf I - \mathbf P_l(t))\mathbf y_l$
     \item  Increment $t=t+1$ and go to step 2 if $l_t<K$, otherwise, set $\hat{\mathcal U}_l = \hat{\mathcal U}_l(t)$

 \end{enumerate}
 \caption{Distributed OMP for sparsity pattern estimation}\label{algo1}
 \end{algorithm}

\subsubsection{Performing step 3 in Algorithm \ref{algo1}}\label{step_3}
To perform step 3 in Algorithm \ref{algo1} we propose the following procedure. \\
\emph{Case I}: Consider the case where the $l$-th node can broadcast its estimated index at each iteration to the rest of the nodes in the  network. i.e. $\mathcal M_l = \bar{\mathcal M} $ where $\bar{\mathcal M}$ contains the indices of all the nodes in the network.   This is a reasonable assumption when there are only a small number of nodes in the network (e.g. cognitive radio networks with a $5-10$ cognitive radios).  During each iteration $t$ of the distributed OMP algorithm, the $l$-th node broadcasts $\lambda_l(t)$.  Consequently, the $l$-th node receives the estimates $\lambda_i(t)$'s for $i\in \bar{\mathcal M}$ from the whole network. Further, let $c(t)$ be a $L$-length vector that contains all the indices estimated at $L$ nodes from step 2  during the $t$-th iteration (i.e. values of  $\lambda_i(t)$ for $i=0,1,\cdots,L-1$).
At $t$-th iteration,  $\lambda_l^*(t)$ is updated as follows:
Check whether there are indices with more than one occurrences (i.e. whether there is any index in the vector $c(t)$ that occurs more than once). If yes, such indices are put in the set $\lambda_l^*(t)$ (such that $\lambda_l^*(t) = \{ set~ of ~indices~ which ~occur ~more~ than~ once \} $. If no, (i.e. there is no index obtained from step 2 such that any two or more nodes agree with, so  that all $L$ indices in $c(t) $ are distinct), then select one index from $c(t)$ randomly and put in  $\lambda_l^*(t)$. In this case, to avoid the same index being selected at subsequent iterations, we force  all nodes to use the same index.

It is noted that when there are approximately equal or more distributed nodes compared to sparsity index $K$, the vector $c(t)$ has at  least one set of two indices with the same value, thus, $\lambda_l^*(t)$ is updated appropriately  most of the  time. Further,  in this case, since each node has the indices received from all the other nodes in the network, every node has the same estimate for $\mathcal U$ after algorithm terminates.


\emph{Case II}: Next, we consider the case where $\mathcal M_l \subset \bar{\mathcal M}$; i.e. each node communicates its estimated index in its neighborhood which has  fewer number of nodes compared to all the nodes in the network. Then, similar to the above case  $\lambda_l^*(t)$ is found based on $c_l(t)$ as the indices which have the most occurrences from $c_l(t)$ where $c_l(t)$ contains the indices received by the $l$-th node from its neighborhood  at $t$-th iteration. However, in this case, since $l$-th node does not receive the estimated indices from the whole network, at each iteration, different nodes may agree upon different indices (however, we see experimentally that for a relatively  large size of $\mathcal M_l$, all the nodes have the same estimated index set at the end). When two neighboring nodes agree upon two different indices at $t$-th iterations, there is a possibility that one node selects the same index at a later iteration than $t$.  To avoid the  $l$-th node selecting  the same index twice, we  perform an additional step; i.e., to check whether  $\lambda_l^*(t)$ determined as above is in $\hat{\mathcal U}_l(t-1)$. If $\lambda_l^*(t)\in \hat{\mathcal U}_l(t-1)$, set $\lambda_l^*(t) = \lambda_l(t)$, otherwise update $\lambda_l^*(t)$ similar to the procedure described in Case I.


\section{Sparse signal detection and sparsity pattern estimation}\label{det_est}
Next, we consider the case where one has to detect whether the sparse signal is present and estimate the sparsity pattern only if the signal is present. We consider the following binary hypothesis testing problem,
\begin{eqnarray}
\mathcal H_1: ~\mathbf y_l&=&\mathbf A_l \mathbf s + \mathbf v_l; ~ \nonumber\\
\mathcal H_0: ~\mathbf y_l&=&\mathbf v_l.\label{hyp2}
\end{eqnarray}
for $l=0,1,\cdots,L-1$ and
where under hypothesis $\mathcal H_1$ the sparse signal is present.
In the following, we extend the  collaborative  algorithm discussed above to first  detect the sparse signal and then to estimate the sparsity pattern without ever reconstructing the signal. Further, we assume that  $L\geq K$.

The idea is based on the properties of the OMP algorithm.
If the signal is not present in the model (\ref{hyp2}), it is very unlikely that two nodes in the network will select the same index of the support set  at any given iteration based on the step 2 in the OMP algorithm presented  in Algorithm \ref{algo1}.  However, when the  signal is present (i.e. hypothesis $\mathcal H_1$), the probability that two nodes select the same index at each iteration is higher as the number of nodes is close to or greater than the sparsity index $K$. That is because, when the signal is present and in relatively high signal-to-noise ratio region, the column index of the projection matrix $\Theta_l$ which is most correlated with the remaining part of the observation can be estimated at the $l$-th node  as  one of the index of  the true sparse support,  and multiple nodes can get the same index during a given iteration.   Taking   this information into account, we extend the algorithm such that it first detects the sparse signal with fewer number of iterations and increase  the number of iterations to find the sparsity pattern only if the signal is detected as described in Algorithm \ref{algo2}.

\begin{algorithm}
At $l$-th node:
\begin{enumerate}
\item Initialize $t=1$, $\hat{\mathcal U}_l(0) = \emptyset $, residual vector $\mathbf r_{l,0} = \mathbf y_l$, $i_{index} = 0$
\item Find the index $\lambda_l(t)$ such that
\begin{eqnarray*}
\lambda_l(t) = \underset{\omega = 1,\cdots,N}{arg~ max} ~ \frac{|<\mathbf r_{l,t-1}, \theta_{l,\omega}>|}{||\mathbf r_{l,t-1}||_2}
\end{eqnarray*}

 \item Update the estimated index set $\lambda_l(t)$ via local communication: $\lambda_l^*(t) = f_l(\lambda_l(t),\{\lambda_{i}(t)\}, i\in \mathcal M_l)$, as discussed in subsection \ref{step_3}
     \item Update $i_{index}$:
     \begin{itemize}
     \item if $unique (c_l(t)) = n(\mathcal M_l)$, $i_{index} = i_{index}+0$
     \item else ( $unique (c_l(t)) < n(\mathcal M_l)$),  $i_{index} = i_{index}+\rho(c_l(t))$
          \end{itemize}as discussed in Subsection \ref{i_index}.
           \item Perform signal detection decision when $t=K_0$
    \begin{itemize}
    \item If $t= K_0$ and $i_{index}(t)\geq I_0$, decide $\mathcal H_1$ and go to step 6. Avoid steps 4 and 5 in subsequent iterations
    \item If $t= K_0$ and  $i_{index}(t)<I_0$ decide $H_0$, set $\hat{\mathcal  U}_l(t) = \emptyset$ and go to step 9
    \end{itemize}
     \item  Set $\hat{\mathcal U}_l(t) = \hat{\mathcal U}_l(t-1) \cup \{\lambda_l^* (t)\}$, and $l_t=n (\hat{\mathcal U}_l(t))$
\item  Compute the projection operator $\mathbf P_l(t) = \Theta_l(\hat{\mathcal U}_l(t)) \left(\Theta_l(\hat{\mathcal U}_l(t)) ^T \Theta_l(\hat{\mathcal U}_l(t)) \right)^{-1} \Theta_l(\hat{\mathcal U}_l(t)) ^T$. Update the residual vector:  $\mathbf r_{l,t} = (\mathbf I - \mathbf P_l(t))\mathbf y_l$

     \item  Increment $t=t+1$ and go to step 2 if $l_t<K$,
     \item set $\hat{\mathcal U}_l = \hat{\mathcal U}_l(t)$

 \end{enumerate}
 \caption{Distributed OMP for sparse signal detection and sparsity pattern estimation}\label{algo2}
 \end{algorithm}
\begin{figure}
  \vspace{0.01cm}
  \includegraphics[scale=0.16,angle=0]{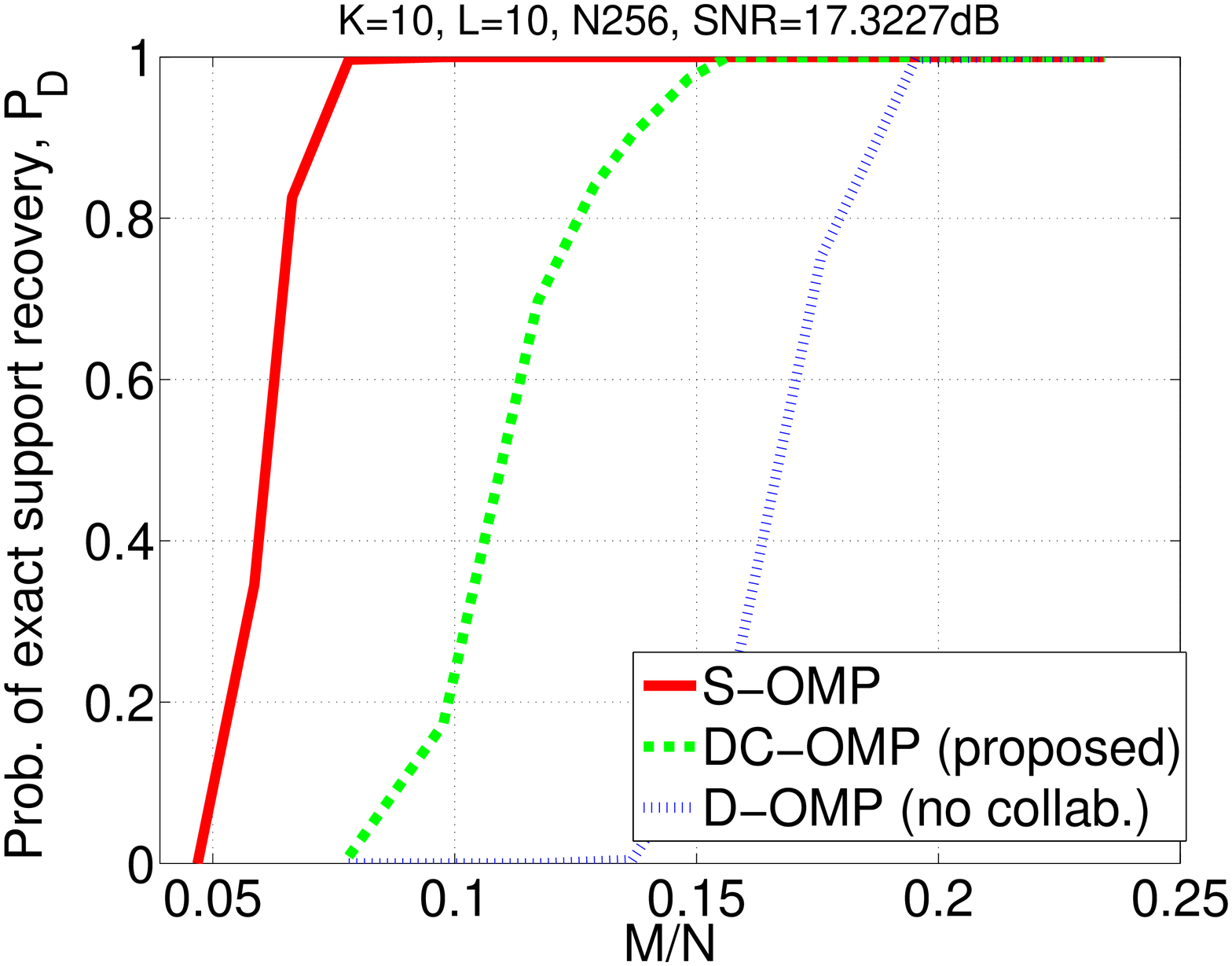}
  \includegraphics[scale=0.16,angle=0]{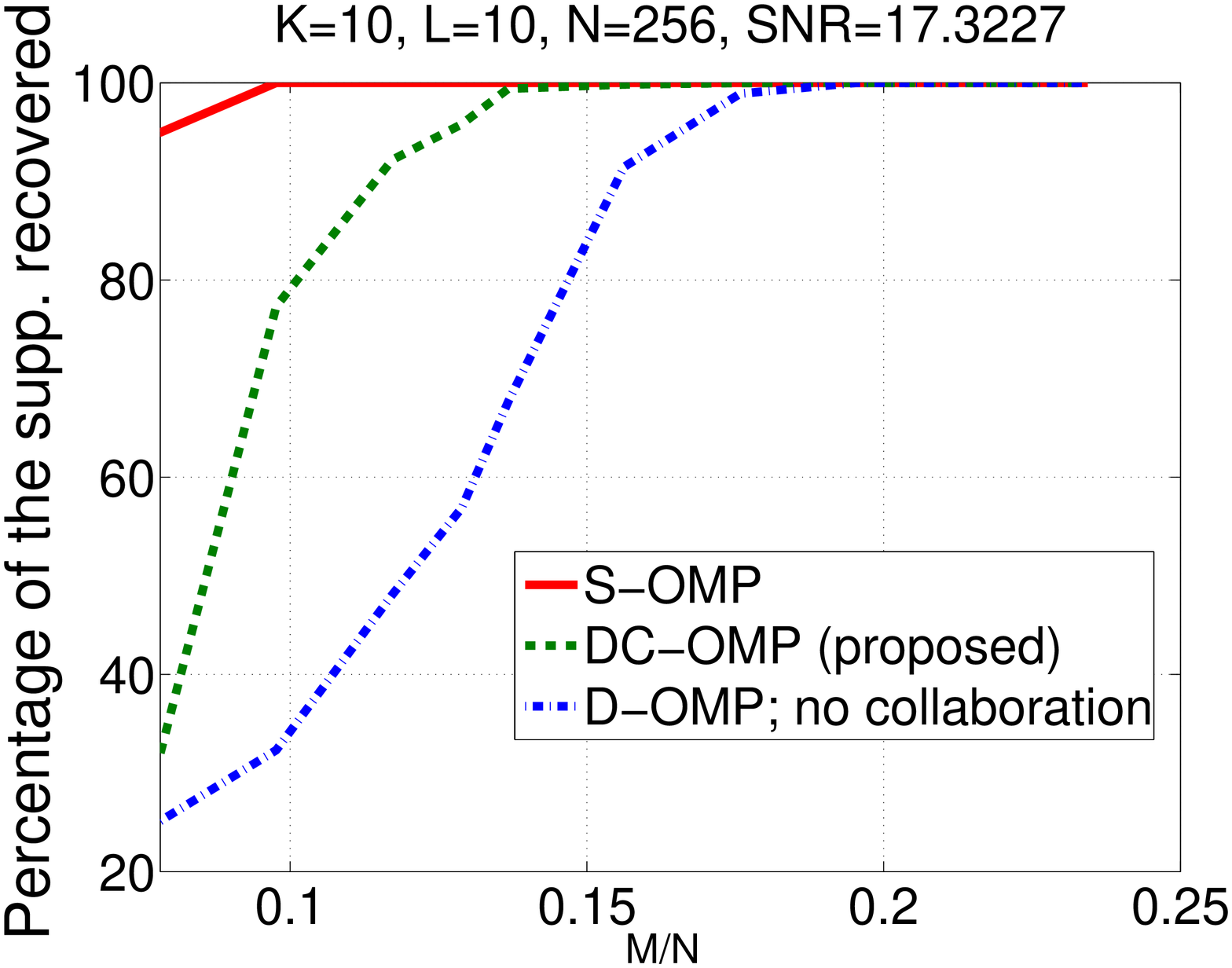}
  \caption{Performance of the sparsity pattern recovery with distributed OMP algorithm: (left) the probability of correctly recovering the sparse support ($P_D = Pr(\hat{\mathbf b} = \mathbf b)$) vs $M/N$ (right) the percentage of the support correctly recovered vs $M/N$; $N=256$, $K=10$, $L=10$, $\bar\gamma = \frac{||\mathbf s||_2^2}{N\sigma_v^2} = 17.3227 dB$}\label{fig_2}
\end{figure}

\subsubsection{Updating $i_{index}$ in step 4} \label{i_index}
Step 4 in Algorithm \ref{algo2} is performed as given below. At $t$-th iteration, $c_l(t)$ contains all the indices received by the  $l$-th node from its neighborhood. The function  $unique (c_l(t))$ gives the number of distinct indices of the support set in  $c(t)$. If all the indices in $c_l(t)$ are different from each other, $unique (c_l(t))$ equals to the number of nodes in the neighborhood of the $l$-th node including itself. If there are any two indices in $c_l(t)$ with the same value, we set the  value of  $\rho(c_l(t)) $ as the number of such indices. After performing $K_0$ (which is less than $K$) number of iterations, if $i_{index}$ in step 4 in Algorithm \ref{algo2} is very small (less than  $I_0$ where $I_0\ll K$), the algorithm decides that no sparse signal is present and terminate the process resulting in the null set as the estimated support set. If $i_{index} \geq I_0$, it decides that the sparse signal is present and continues estimating the support set similar to  Algorithm \ref{algo1}.

\section{Simulation Results}\label{numerical}
In this section, we present  some simulation results to demonstrate the performance of proposed  algorithms for sparsity pattern  recovery based on distributed CS based measurements. In the following, we assume that the entries of each projection matrix $\mathbf A_l$ for $l=0,1,\cdots, L-1$ are drawn from a Gaussian ensemble with mean zero and  variance $\frac{1}{N}$.
\begin{figure}
\centerline{\epsfig{figure=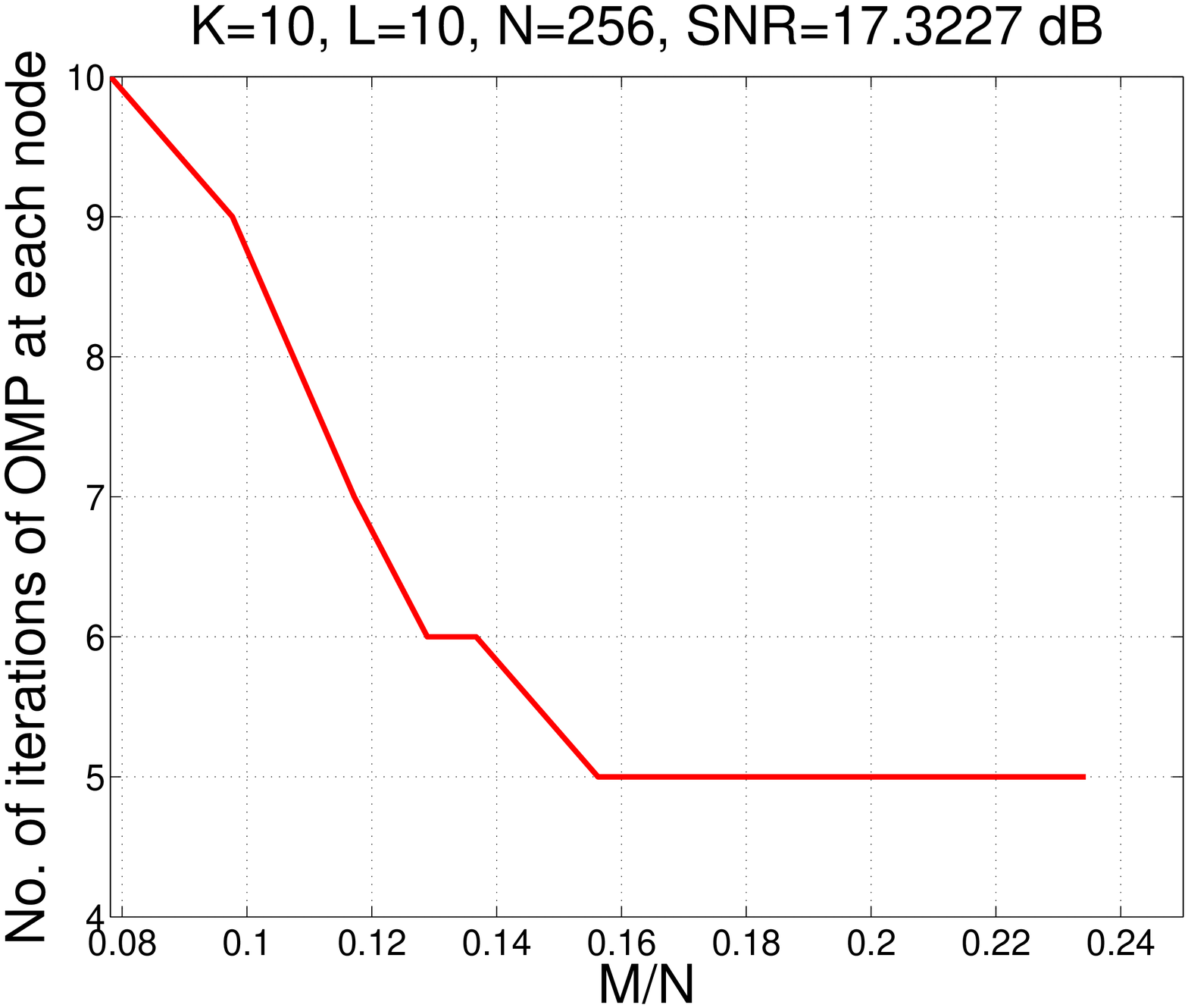,width=5.5cm}}
\caption{Number of iterations of OMP at each node with the proposed Algorithm  \ref{algo1} vs $M/N$; $N=256$, $K=10$, $L=10$, $\bar\gamma = \frac{||\mathbf s||_2^2}{N\sigma_v^2} = 17.3227 dB$, }
  \label{fig_3}
\end{figure}

To compare the performance of the proposed  Algorithm \ref{algo1} with other approaches, we consider two benchmark cases. (i). Distributed OMP with no collaboration: in this case, each node performs OMP independently  to obtain the support set estimate $\hat{\mathcal U}_l(t)$,  i.e., the step 3 is eliminated in Algorithm \ref{algo1} and set $\lambda_l^*(t) = \lambda_l(t)$. To fuse the support sets estimated at individual nodes,  each node transmits its estimated support to a fusion center and performs a majority rule based fusion scheme to obtain a global estimate of $\mathcal U$.
(ii). Simultaneous OMP (S-OMP) \cite{Tropp2}: S-OMP algorithm is carried out using  all the raw observations at the fusion center. Thus, this scheme requires  each node to transmit its $M$-length observation vector as well as the projection matrix $\mathbf A_l$ to the fusion center.

In Figures \ref{fig_2} and \ref{fig_3}, we illustrate the  performance of the sparsity pattern estimation based on proposed DC-OMP  as in Algorithm \ref{algo1} in terms of different  performance metrics. In both  figures, we  let  $K=10$, $N=256$,  $L=10$, and SNR at each node, defined as $\bar\gamma = \frac{||\mathbf s||_2^2}{N \sigma_v^2} = 17.3227 dB$. Also, in Figures \ref{fig_2} and \ref{fig_3} we  assume that $\mathcal M_l = \bar{\mathcal M} $ as considered in case I in Subsection \ref{step_3}. Then the estimated support set at each node based on Algorithm \ref{algo1} is the same. In Fig. \ref{fig_2}, by performing $10^4$ runs and averaging over 10 trials, we plot the probability of correctly recovering the full  support set, $P_D= Pr(\hat{\mathbf b} =\mathbf b )$ (left) and the percentage of the support set that is estimated correctly (right) vs $M/N$ where $M$ is the number of compressive measurements at each node. It can be seen from Fig. \ref{fig_2} that, at relatively small values of $M/N$, the proposed algorithm outperforms  D-OMP with no collaboration. In resource constrained distributed networks, especially in sensor networks, it is desirable to perform the desired task by employing  less measurement data  (i.e. with small $M$) at each node distributing the computational complexity among nodes to save the overall node power. Thus, fusing the estimated indices of the non-zero coefficients at each iteration of the OMP algorithm ensures a higher performance in  exact sparsity pattern recovery compared to that when   OMP is performed at each node independently. However, as $M/N$ increases, $Pr(\hat{\mathbf b} =\mathbf b )$ of both algorithms converge to 1, since when the number of compressive measurement at each node increases,  OMP (with or without collaboration) works better and recovers the sparsity pattern almost exactly with a single measurement vector. It has been shown in \cite{tropp1,Fletcher2} that OMP requires approximately $M\geq (1+\epsilon) 4K\log N$, $\epsilon>0$  measurements for reliable sparsity pattern recovery in the noise free case. Thus, even with a very small $M$ at each node, having $L$ number of nodes,  S-OMP achieves this limit and provides a significant performance gain compared to the proposed algorithm at very small $M/N$ values. However, S-OMP requires a considerable communication overhead compared to the proposed algorithm. Further, in the proposed algorithm, each node has the same estimator at the end in contrast to the  centralized  S-OMP.
\begin{figure}
  \vspace{0.01cm}
  \includegraphics[scale=0.16,angle=0]{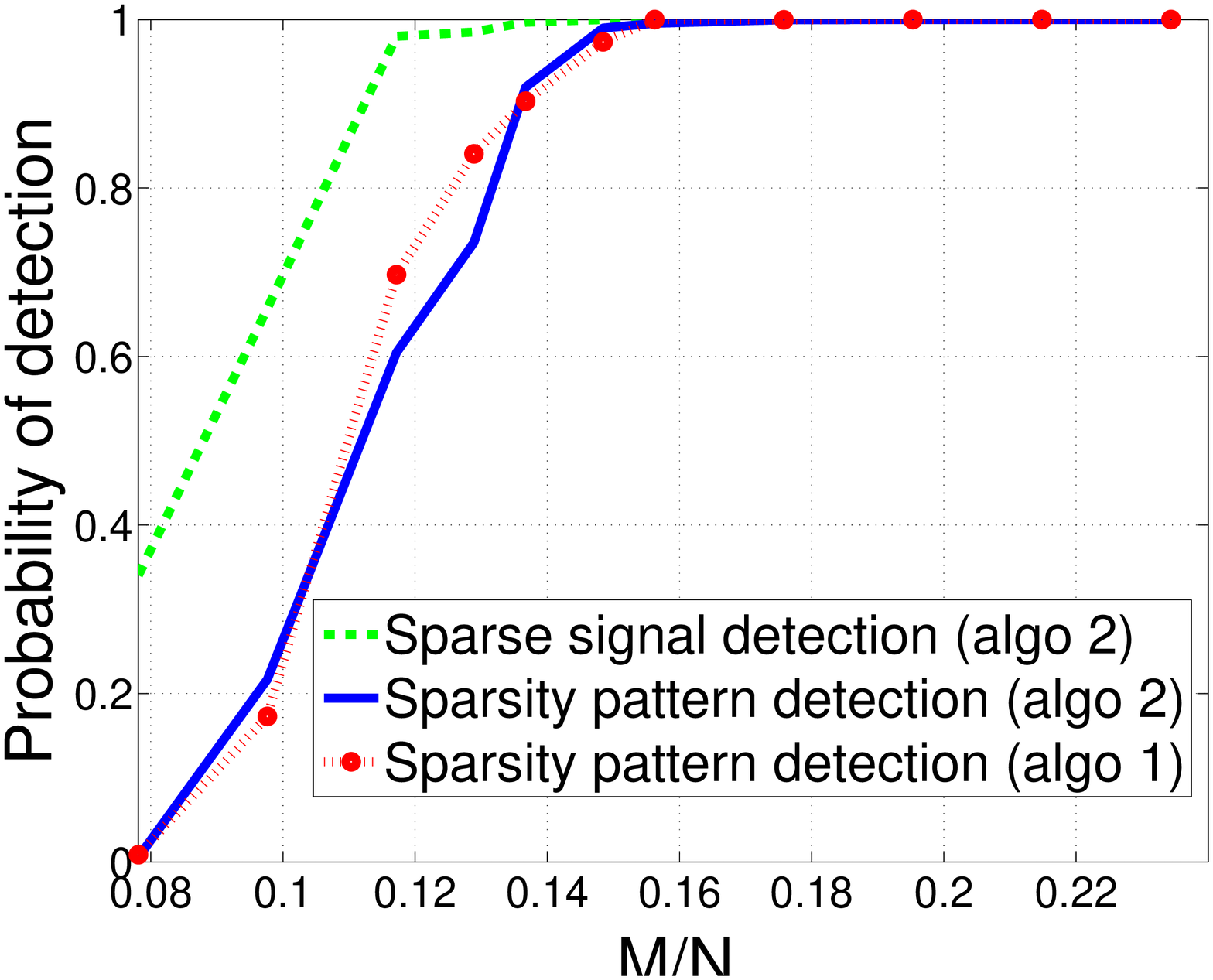}
  \includegraphics[scale=0.16,angle=0]{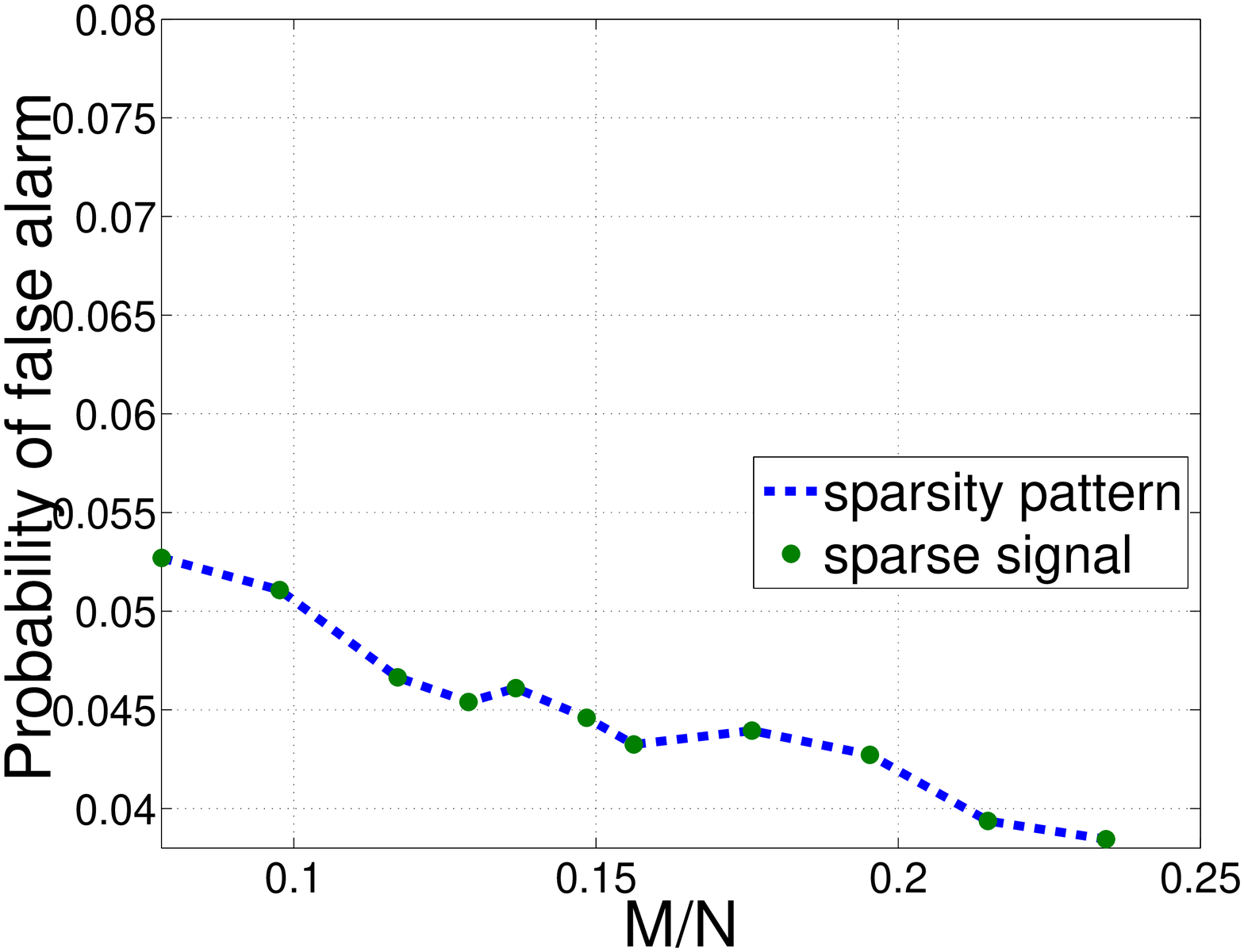}
  \caption{Performance of the sparse signal detection and sparsity pattern recovery with Algorithm \ref{algo2}; $N=256$, $K=10$, $L=10$, $\bar\gamma = \frac{||\mathbf s||_2^2}{N\sigma_v^2} = 17.3227 dB$; (left) Probability of detection, (right) probability of false alarms}\label{fig_4}
\end{figure}
In Fig. \ref{fig_2}, we further plot the percentage of support that is correctly recovered. for example, at $\frac{M}{N}\approx 0.1$, the proposed algorithm correctly recovers approximately $75 \%$  of the  support while D-OMP with no collaboration recovers only about $30 \%$ of the support. Thus, from both sub figures of Fig. \ref{fig_2}, significant  performance gain  is observed  via the proposed algorithm compared to D-OMP with no collaboration with small $M$, which is the more desirable scenario in resource constrained distributed networks. To further illustrate the efficiency of the proposed algorithm, in Fig. \ref{fig_3}, we plot the number of iterations of the DC-OMP algorithm that each node has to  perform in recovering the sparsity pattern. It is observed from Fig. \ref{fig_3} that as $M/N$ increases, the proposed algorithm estimates the sparsity pattern reliably by executing only $\approx K/2$ number of  iterations at each node. When $M/N$ increases, as observed from Fig. \ref{fig_2}, the performance of both DC-OMP and D-OMP with no collaboration converges to the same level  but DC-OMP requires very small number of iterations at each node to achieve that performance compared to that with D-OMP with no collaboration which requires $K$ number of iterations at each node irrespective of the value of $M/N$.

In Fig. \ref{fig_4}, we illustrate the performance of Algorithm \ref{algo2} for detecting the sparse signal before estimating the sparsity pattern. We plot the performance of sparse signal detection as well as the sparsity pattern estimation in Fig. \ref{fig_4}. For sparse signal detection, probability of detection and the false alarm are given by  $P_D^{s} = Prob(\delta = 1|\mathcal H_1)$ and $P_F^s = Prob(\delta = 1|\mathcal H_0)$, respectively where $\delta$ is the detection decision. For sparsity pattern detection, the probability of detection is given by  $P_D^u = Prob(\hat{\mathbf b} = \mathbf b^1  | \mathcal H_1)Prob(\mathcal H_1) + Prob(\hat{\mathbf b} = \mathbf b^0  | \mathcal H_0)Prob(\mathcal H_0)$, where we redefine the variables such that $\mathbf b^1$ is the binary support of the signal $\mathbf s$ (i.e. the support under $\mathcal H_1$) while $\mathbf b^0$ denotes the vector with all zeros (binary support under $\mathcal H_0$). Similarly the probability of false alarm, is given by   $P_F^u = Prob(\hat{\mathbf b} = \mathbf b^1  | \mathcal H_0)Prob(\mathcal H_1) + Prob(\hat{\mathbf b} = \mathbf b^0  | \mathcal H_1)Prob(\mathcal H_0)$. Further, in Fig. \ref{fig_4} we use the same values for the parameters $N, K, L$ and $\bar\gamma$ as used in Figures \ref{fig_2} and \ref{fig_3} and $K_0=3$ and $I_0=2$.  From Fig. \ref{fig_4}, it is seen  that Algorithm \ref{algo2} reliably detects the sparse signal even with a very small value of $M/N$, and the performance of the sparsity pattern recovery after detecting the signal  has  performance that is close  to that when  the sparsity pattern recovery is done as in Algorithm \ref{algo1} (where it is known \emph{a priori} that the signal is present).

\section{conclusion}\label{conclusions}
In this paper, we addressed  the problem of recovering a common  sparsity pattern   based on CS measurement vectors collected at distributed nodes in a distributed network. A distributed greedy algorithm based on OMP is proposed to estimate the sparsity pattern via collaboration in which each distributed node is required to perform  less number of iterations of the greedy algorithm compared to the sparsity index.  When it is not known \emph{a priori} that a sparse signal is present or not, the algorithm was extended to perform detection of the sparse signal with a fewer number of iterations before completely recovering the sparsity pattern. The proposed algorithm is shown to have significant performance gains compared to that with each node performing OMP independently and then fusing  the estimated supports to achieve a global estimate. Complete theoretical analysis of the algorithm will be considered in a future work.
\newpage
\bibliographystyle{IEEEtran}
\bibliography{IEEEabrv,bib1}
\end{document}